# Three-Body Baryonic $\bar{B}^0 \to \Lambda_c^+ \bar{p} \pi^0$ Decay

S. Rafibakhsh*, H. Mehraban†

Physics Department, Semnan University

P.O.Box 35195-363, Semnan, Iran

**Abstract**

We study the three-body baryonic decay $\bar{B}^0 \to \Lambda_c^+ \bar{p} \pi^0$ based on the factorization approach. The most important Feynman diagrams for this decay mode are factorizable and non-factorizable contributions, so that the former plays the key role in the decay amplitude and the latter affects the coefficient $a_2$ . We calculate the decay branching ratio for some values of $a_2$ in the range of $0.16-0.40$ and we find that the value of $a_2 = 0.17$ leads us to the result of $(1.50 \pm 0.21) \times 10^4$ which evidently agrees with the experimental result $(1.55 \pm 0.17 \pm 0.08) \times 10^4$ reported by BABAR.

## I. Introduction

Due to large B meson mass, its decays to vast numbers of hadrons even baryons, which in fact cover 7% of all B meson decays, are possible [1, 2]. The first baryonic B decays i.e. three- and four-body modes, were observed in the late 1980s and began to be studied in the early 1990s [3]. The first two-body baryonic mode $\bar{B}^0 \to \Lambda_c^+ \bar{p}$ was observed around two decades later, in 2002 [4, 5]. Furthermore, the number of the observed two-body baryonic decays is much less than the multi-body modes. This is due to the fact that the configuration of a baryon pair accompanied by at least a single meson in the final state, is more favorable by the B meson as the fast recoil meson carries away a considerable amount of the released energy and this results in reducing the effective mass of the baryon pair and hence the pair is easier to produce [3, 6].

* rafibakhsh@semnan.ac.ir

† hmehraban@semnan.ac.ir

Theoretical studies and calculation of the branching ratio of the observed decays and predicting the branching ratio of the non-observed modes, by means of different models have ever been of great interest in phenomenology.

In the present paper, we have calculated the branching ratio of the three-body B decay $\bar{B}^0 \to \Lambda_c^+ \bar{p} \pi^0$ using factorization hypothesis whose notable success in computing and predicting branching ratios has been proven in recent years.

## II. Formalism

To compute the branching ratio of any decays, the first step is depicting the relevant Feynman diagrams. The quark diagrams for the decay $\bar{B}^0 \to \Lambda_c^+ \bar{p} \pi^0$ are demonstrated in Fig. 1. This decay receives a factorizable internal W-emission contribution (Fig. 1(a)), two non-factorizable internal W-emission contributions (Figs. 1(b)-1(c)) and a W-exchange mode (Fig. 1(d)). Among these contributions, as pointed out in [7] the W-exchange mode is insignificant and hence can be neglected. Moreover the non-factorizable effects are not calculable in practice, however we will return to them and go through their affects later. Hence we can simply focus on the factorizable color-suppressed internal W-emission contribution in which the baryon pair is current-produced.

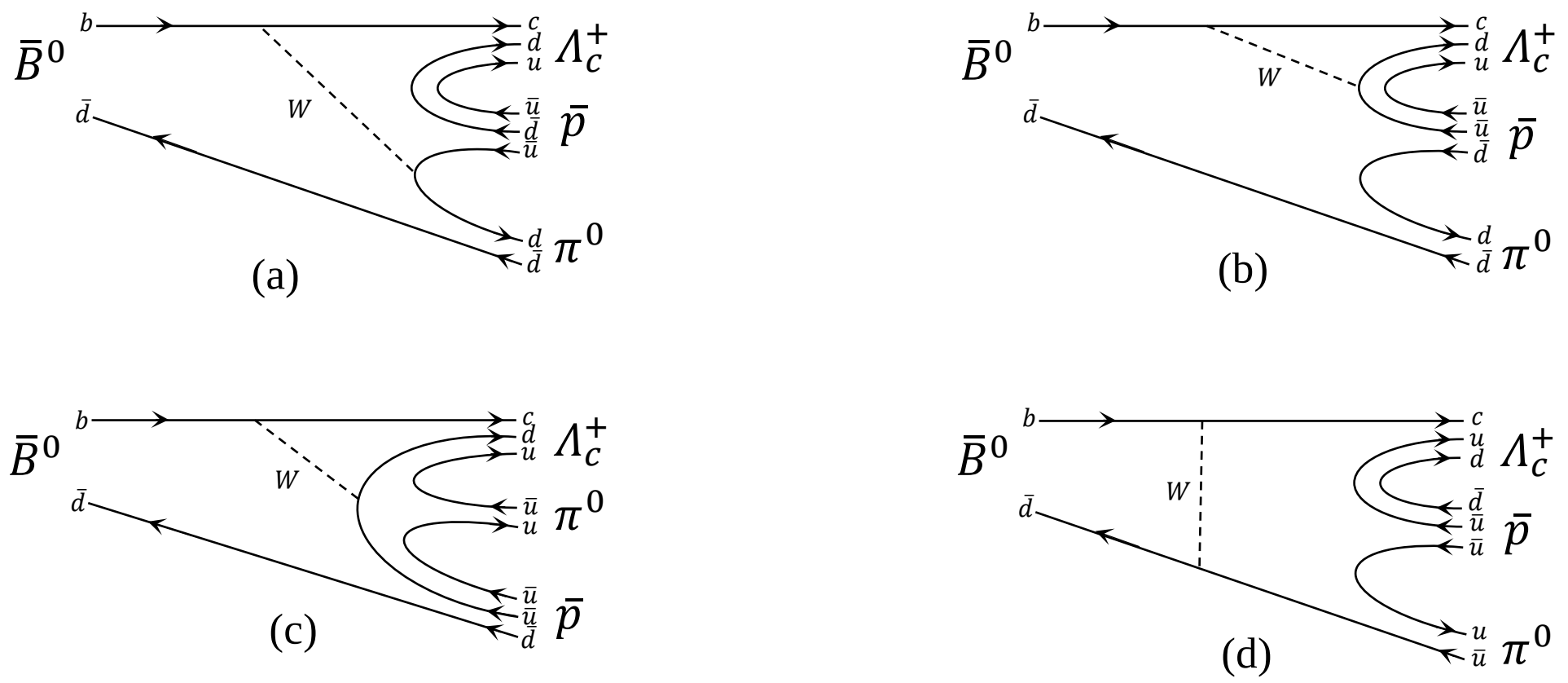

Fig. 1. Quark diagrams for the decay $\bar{B}^0 \to \Lambda_c^+ \bar{p} \pi^0$. (a) factorizable internal W-emission, (b,c) non-factorizable internal W-emission, (d) W-exchange.

Based on the factorization assumption, decay amplitudes are approximately factorized into hadronic matrix elements multiplied by coefficients $a_i$ derived from the Wilson coefficients [8-10]. Thus the amplitude of the decay $\bar{B}^0 \to \Lambda_c^+ \bar{p} \pi^0$ in terms of the effective Hamiltonian at quark level under factorization framework, reads

$$\mathcal{M} = \frac{G_F}{\sqrt{2}} V_{cb} V_{ud}^* a_2 \left\langle \bar{B}^0 \middle| \left(\bar{d}b\right)_{V-A} \middle| \pi^0 \right\rangle \langle \bar{p}\Lambda_c^+ | (\bar{c}u)_{V-A} | 0 \rangle \,, \tag{1}$$

where $G_F$ is the Fermi constant, $V_{ij}$ are the CKM matrix elements, $(\bar{q}_1 q_2)_{V(A)}$ stands for $\bar{q}_1 \gamma_\mu (\gamma_5) q_2$ and the parameter $a_2 = c_2^{eff} + c_1^{eff}/N_c^{eff}$ with the effective Wilson coefficients $c_i^{eff}$ and the effective color number $N_c^{eff}$ ranging between 2 and ∞, will be specified later.

The matrix element of the transition $B \to \pi$ has the expression [11]

$$\begin{aligned} \left\langle \bar{B}^0 \middle| \left(\bar{d}b\right)_{V-A} \middle| \pi^0 \right\rangle &= \langle \bar{B}^0 | \bar{d}(1-\gamma^5)\gamma^\mu b | \pi^0 \rangle \\ &= (p_B + p_{\pi^0})^\mu F_1^{B\pi^0}(t) + \left( \frac{{m_B}^2 - {m_{\pi^0}}^2}{t} \right) q^\mu \left( F_0^{B\pi^0}(t) - F_1^{B\pi^0}(t) \right), \end{aligned} \tag{2}$$

where $q_\mu = (p_B - p_\pi)_\mu = \left(p_{\Lambda_c} + p_p\right)_\mu$, $t \equiv q^2$ and $m_i$s are the mass of the hadrons. There are various models to determine the form factors. In our calculation, the relativistic covariant light-front quark model is applied to evaluate the mesonic form factors given by [11-13]

$$F_1^{B\pi^0}(t) = \frac{F_1^{B\pi^0}(0)}{\left(1 - \frac{t}{{m_V}^2}\right)\left(1 - \sigma_{11}\frac{t}{{m_V}^2} + \sigma_{12}\frac{t^2}{{m_V}^4}\right)}, \tag{3}$$

$$F_0^{B\pi^0}(t) = \frac{F_0^{B\pi^0}(0)}{1 - \sigma_{01}\frac{t}{{m_V}^2} + \sigma_{02}\frac{t^2}{{m_V}^4}}. \tag{4}$$

Moreover, the baryonic matrix element is parametrized as [14, 15]

$$\begin{aligned} \langle \bar{p}\Lambda_c^+ | (\bar{c}u)_{V-A} | 0 \rangle &= \langle \bar{p} | (\bar{c}u)_{V-A} | \Lambda_c^+ \rangle \\ &= \bar{u}_{\Lambda_c^+}(p_1) \Bigg[ f_1(q^2)\gamma_\mu + i\frac{f_2(q^2)}{m_{\Lambda_c^+} + m_p}\sigma_{\mu\nu}q^\nu \\ &\quad - \left( g_1(q^2)\gamma_\mu + i\frac{g_2(q^2)}{m_{\Lambda_c^+} + m_p}\sigma_{\mu\nu}q^\nu + \frac{g_3(q^2)}{m_{\Lambda_c^+} + m_p}q_\mu \right)\gamma_5 \end{aligned}$$

$$+\frac{f_3(q^2)}{m_{\Lambda_c^+}+m_p}q_\mu\Bigg]v_{\bar{p}}(p_2)\,, \tag{5}$$

where $f_i$ and $g_i$ are the baryonic form factors in the non-relativistic quark model at $q=0$. In practice, these form factors cannot be estimated directly utilizing the above-mentioned model. The alternative approach is the pole model which is customarily applied to evaluate the $q^2$ dependence of the form factors [15] [see Fig. 2]:

$$f_i(q^2)=f_i(q_m^2)\left(\frac{1-q_m^2/m_{D^{*0}}^2}{1-q^2/m_{D^{*0}}^2}\right)^n,$$
$$g_i(q^2)=g_i(q_m^2)\left(\frac{1-q_m^2/m_{D^{*0}_2}^2}{1-q^2/m_{D^{*0}_2}^2}\right)^n, \tag{6}$$

with $q_m=\left(m_{\Lambda_c}-m_p\right)^2$ and as the transition is heavy-to-light, the dipole dependence of the form factor is feasible. Thus the parameter $n$ is taken to be equal to 2 [14, 15].

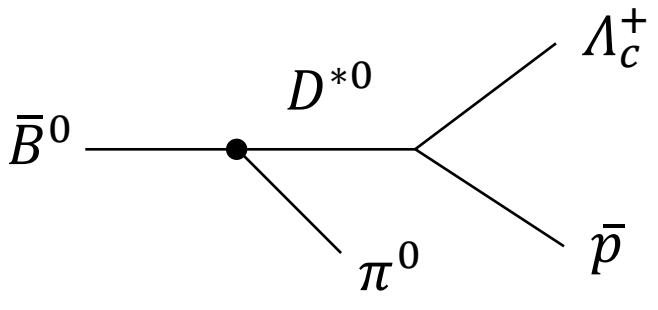

Fig. 2. The relevant pole diagram of the Fig. 1(a).

Finally, the decay amplitude is obtained as follow

$$\mathcal{M}=\frac{G_F}{\sqrt{2}}V_{cb}V_{ud}^*a_2\big[(A+B\gamma_5)\not{p}_\pi+(C\gamma_5+D)\big]\,, \tag{7}$$

where

$$A=2f_1^{\Lambda_c p}(t)F_1^{B\pi}(t)+2f_2^{\Lambda_c p}(t)F_1^{B\pi}(t)\left(m_{\Lambda_c}+m_p\right)/m_{\Lambda_c}\,,$$

$$B=-2g_1^{\Lambda_c p}(t)F_1^{B\pi}(t)-2g_2^{\Lambda_c p}(t)F_1^{B\pi}(t)\left(m_{\Lambda_c}-m_p\right)/m_{\Lambda_c}\,,$$

$$C=-\left(m_{\Lambda_c}+m_p\right)g_1^{\Lambda_c p}(t)\times\left[F_1^{B\pi}(t)+\left(F_0^{B\pi}(t)-F_1^{B\pi}(t)\right)(m_B^2-m_\pi^2)/t\right]$$

$$+2g_2^{\Lambda_c p}(t)F_1^{B\pi}(t)\left(p_{\Lambda_c}-p_p\right).p_\pi/m_{\Lambda_c}-g_3^{\Lambda_c p}(t)F_0^{B\pi}(t)\left(m_B^2-m_\pi^2\right)/m_{\Lambda_c}\,,$$

$$D=\left(m_{\Lambda_c}+m_p\right)f_1^{\Lambda_c p}(t)\times\left[F_1^{B\pi}(t)+\left(F_0^{B\pi}(t)-F_1^{B\pi}(t)\right)\left(m_B^2-m_\pi^2\right)/t\right]$$

$$+2f_2^{\Lambda_c p}(t)F_1^{B\pi}(t)\left(p_{\Lambda_c}-p_p\right).p_\pi/m_{\Lambda_c}+f_3^{\Lambda_c p}(t)F_0^{B\pi}(t)\left(-m_\pi^2\right)/m_{\Lambda_c}\,. \tag{8}$$

Then using the Casimir trick for every single one of the 16 terms of the Eq. (8), the amplitude's absolute square is obtained

$$\sum|\mathcal{M}|^2=G_F^2\{|A|^2\left[\left(m_B^2+m_p^2-t-s\right)\left(s-m_p^2-m_\pi^2\right)-m_\pi^2\left(t-\left(m_{\Lambda_c}-m_p\right)^2\right)\right]$$

$$+2Re(AD^*)\left[m_{\Lambda_c}\left(s^2-{m_p}^2-{m_\pi}^2\right)-m_p\left({m_B}^2+{m_p}^2-t-s\right)\right]\}\,, \tag{9}$$

where $s$ and $t$ are the invariants in the Dalitz plot analysis defined by [16]

$$s=\left(p_p+p_\pi\right)^2=\left(p_B-p_{\Lambda_c}\right)^2,\qquad t=\left(p_{\Lambda_c}+p_p\right)^2=\left(p_B-p_\pi\right)^2. \tag{10}$$

with $p_i$s being the four-momentum of the hadrons.

The decay width for a three-body decay is given by

$$\Gamma=\frac{1}{(2\pi)^3}\frac{1}{32{m_B}^3}\int_{s_{min}}^{s_{max}}\int_{t_{min}}^{t_{max}}\left(\sum|\mathcal{M}|^2\right)dsdt\,, \tag{11}$$

where the boundaries of the phase space have taken the standard form used in the Dalitz plot as following

$$s_{min}=\left(m_p+m_\pi\right)^2,\qquad s_{max}=\left(m_B-m_{\Lambda_c}\right)^2 \tag{12}$$

and

$$t_{\substack{min\\max}}={m_{\Lambda_c}}^2+{m_p}^2-\frac{1}{s}[(s-{m_B}^2+{m_\pi}^2)\left(s+{m_{\Lambda_c}}^2-{m_p}^2\right)$$

$$\mp\lambda^{1/2}(s,{m_B}^2,{m_\pi}^2)\lambda^{1/2}\left(s,{m_{\Lambda_c}}^2,{m_p}^2\right)]\,, \tag{13}$$

with

$$\lambda(a,b,c) = a^2 + b^2 + c^2 - 2(ab + ac + bc).$$

Ultimately, the branching ratio is calculated by

$$\mathrm{BR} = \frac{\Gamma(\bar{B}^0 \to \Lambda_c^+ \bar{p}\pi^0)}{\Gamma_{tot}}. \quad (14)$$

## III. Numerical Results

For the numerical analysis, the CKM matrix elements and the hadron masses are given by [17]

$$V_{cb} = 42.2 \times 10^{-3} \pm 8 \times 10^{-4}, \quad V_{ud} = 0.97 \pm 2.1 \times 10^{-4},$$
$$m_{B^0} = 5279.55 \pm 0.17\, MeV, \quad m_{\Lambda_c} = 2286.46 \pm 0.14\, MeV,$$
$$m_p = 0.938\, MeV, \quad m_{\pi^0} = 134.9770 \pm 0.0005\, MeV$$

and the pole masses in Eq. (6) have the following quantities [16]

$$m_{D^{*0}} = 2006 \pm 0.5\, MeV, \quad m_{D_2^{*0}} = 2460 \pm 0.4\, MeV.$$

In addition, the required parameters to calculate the mesonic form factors in Eq. (3) and Eq. (4) are taken to be [11]

$$m_V = 5.32\, GeV, \quad F_{1,0}^{B\pi^0}(0) = 0.29,$$
$$(\sigma_{11}, \sigma_{12}) = (0.48, 0), \quad (\sigma_{01}, \sigma_{02}) = (0.76, 0.28),$$

and we also need to know the parameters $f_i(q_m^2)$ and $g_i(q_m^2)$ to evaluate the baryonic form factors in Eq. (5) [15]

$$f_1(q_m^2) = g_1(q_m^2) = 0.80, \quad f_2(q_m^2) = g_2(q_m^2) = -0.21,$$
$$f_3(q_m^2) = g_3(q_m^2) = -0.07.$$

Furthermore, in the generalized factorization hypothesis, the parameter $a_2$ is greatly affected by the non-factorizable terms. As this parameter is not universal, it is extracted from the data [15] and can be assumed as a free parameter ranging between 0.16 and 0.40 [9,10]. Thus we calculate the decay branching ratio, using different values of $a_2$. According to our results, listed in Table 1, it is obvious that the branching ratio increases by $a_2$ and larger values of $a_2$ result in larger values of branching ratios which exceed the experimental branching ratio substantially. Thus we ignore larger values and we find that the value of $a_2 = 0.17 \pm 0.03$ leads us to a result which is more consistent with the experimental result reported by BABAR

$$BR_{EXP}(\bar{B}^0 \to \Lambda_c^+ \bar{p} \pi^0) = (1.55 \pm 0.17 \pm 0.08) \times 10^{-4}.$$

TABLE 1. Branching ratios of the decay $\bar{B}^0 \to \Lambda_c^+ \bar{p} \pi^0$ (in units of $10^{-4}$) for some values of $a_2$ .

| $a_2$ | BR |
|---|---|
| $0.16 \pm 0.04$ [9] | $1.33 \pm 0.25$ |
| $0.17 \pm 0.03$ [6] | $1.50 \pm 0.21$ |
| $0.20 \pm 0.05$ [9] | $2.08 \pm 0.32$ |
| $0.22 \pm 0.06$ [9] | $3.51 \pm 0.37$ |

## IV. Conclusion

In this work, we have studied the three-body baryonic decay $\bar{B}^0 \to \Lambda_c^+ \bar{p} \pi^0$ within the framework of the generalized factorization approach. According to the Feynman diagrams we have depicted, the decay involves factorizable and non-factorizable contributions in addition to a w-exchange mode. The factorizable mode is dominant and the non-factorizable contributions affect the amount of the coefficient $a_2$ in the decay amplitude. The mesonic and baryonic form factors are determined applying covariant light-front quark model and non-relativistic quark model, respectively. Since the baryonic form factors are not calculable directly, the pole model approach is utilized instead. We have calculated the decay branching ratio for some value of $a_2$ and we find that the value of of $a_2 = 0.17$ leads us to a result which seems to be reasonably consistent with the experimental result reported by BABAR.